\documentclass[fleqn,usenatbib]{mnras}

\usepackage{newtxtext,newtxmath}
\usepackage[T1]{fontenc}

\DeclareRobustCommand{\VAN}[3]{#2}
\let\VANthebibliography\thebibliography
\def\thebibliography{\DeclareRobustCommand{\VAN}[3]{##3}\VANthebibliography}

\usepackage{graphicx}	% Including figure files
\usepackage{amsmath}	% Advanced maths commands
\usepackage[version=4]{mhchem}
\usepackage{hyperref}
\usepackage[separate-uncertainty=true]{siunitx}
\usepackage{bm}
\usepackage{mismath}
\usepackage{float}

\DeclareSIUnit\bar{bar}
\DeclareSIUnit\AU{AU}
\DeclareSIUnit\dex{dex}
\DeclareSIUnit\erg{erg}
\DeclareSIUnit\day{day}
\DeclareSIUnit\year{yr}
\newcommand{\WPMS}{\watt\per\meter\squared}

\title[Convective shutdown on lava worlds]{Convective shutdown in the atmospheres of lava worlds}

\author[H. Nicholls et al.]{
Harrison Nicholls,$^{1}$\thanks{E-mail: harrison.nicholls@physics.ox.ac.uk}
Raymond T. Pierrehumbert,$^{1}$
Tim Lichtenberg,$^{2}$
Laurent Soucasse,$^{3}$
Stef Smeets$^{3}$
\\
$^{1}$Atmospheric, Oceanic, and Planetary Physics, Department of Physics, University of Oxford, Oxford OX1 3PU, United Kingdom
\\
$^{2}$Kapteyn Astronomical Institute, University of Groningen, P.O. Box 800, 9700 AV Groningen, The Netherlands
\\
$^{3}$Netherlands eScience Center, Science Park 402, 1098 XH Amsterdam, The Netherlands
}

\date{Accepted XXX. Received YYY; in original form ZZZ}
\pubyear{\the\year{}}

\begin{document}
\label{firstpage}
\pagerange{\pageref{firstpage}--\pageref{lastpage}}
\maketitle

% Abstract of the paper (<250 words, no references)
\begin{abstract}
Atmospheric energy transport is central to the cooling of primordial magma oceans. Theoretical studies of atmospheres on lava planets have assumed that convection is the only process involved in setting the atmospheric temperature structure. This significantly influences the ability for a magma ocean to cool. It has been suggested that convective stability in these atmospheres could preclude permanent magma oceans.
We develop a new 1D radiative-convective model in order to investigate when the atmospheres overlying magma oceans are convectively stable. Using a coupled interior-atmosphere framework, we simulate the early evolution of two terrestrial-mass exoplanets: TRAPPIST-1 c and HD 63433 d.
Our simulations suggest that the atmosphere of HD 63433 d exhibits deep isothermal layers which are convectively stable. However, it is able to maintain a permanent magma ocean and an atmosphere depleted in \ce{H2O}. It is possible to maintain permanent magma oceans underneath atmospheres without convection. Absorption features of \ce{CO2} and \ce{SO2} within synthetic emission spectra are associated with mantle redox state, meaning that future observations of HD 63433 d may provide constraints on the geochemical properties of a magma ocean analogous with the early Earth. Simulations of TRAPPIST-1 c indicate that it is expected to have solidified within \SI{100}{\mega\year}, outgassing a thick atmosphere in the process. Cool isothermal stratospheres generated by low molecular-weight atmospheres can mimic the emission of an atmosphere-less body. 
Future work should consider how atmospheric escape and chemistry modulates the lifetime of magma oceans, and the role of tidal heating in sustaining atmospheric convection.

\end{abstract}


%%%%%%%%%%%%%%%%% BODY OF PAPER %%%%%%%%%%%%%%%%%%

\section{Introduction}
\label{sec:intro}

Whether formed by accretion, giant impacts, or through other processes, magma oceans are thought to be common to the early stages of planetary evolution \citep{abe_early_1986, rubie_formation_2007, salvador_venus_2023}. These magma oceans may then be sustained by continuous energy delivery from stellar irradiation or by tidal heating \citep{ demory_55cnc_2016, hay_tides_2019}. Outgassing of volatiles, such as \ce{H2O} and \ce{CO2}, from magma oceans has been shown to prolonging their lifetimes due to the induction of a greenhouse effect \citep{abe_early_1986, hamano_lifetime_2015, nicholls_proteus_2025}. In equilibrium with an underlying permanent magma ocean, the composition of an overlying atmosphere is strongly controlled by the geochemical (e.g. redox state) and physical conditions of the magma ocean \citep{nicholls_proteus_2025, gaillard_redox_2022}. Therefore, characterising or constraining the composition of degassed atmospheres can provide insight into the geochemistry of exoplanetary interiors \citep{kite_exchange_2016, baumeister_redox_2023}. For planets which eventually do solidify, the evolutionary outcome of these early magma oceans is important for setting the initial conditions for solid-state evolution, potential plate tectonics, and the reservoir of volatiles in the atmosphere \citep{foley_plate_2014, baumeister_redox_2023, sossi_redox_2020}. The physical interaction between the outgassing of radiatively opaque volatiles and the secular cooling (and associated degassing) from an underlying magma ocean also necessitates the use of evolutionary calculations to link the observed state of planets with permanent magma oceans (lava planets) to the conditions set by accretion \citep{elkins_linked_2008, lichtenberg_redox_2021}. For planets with permanent magma oceans, the favourable dissolution of volatiles (particularly \ce{H2O}) into the melt could mean that atmosphere-interior coupling acts to buffer the atmosphere against escape, thereby prolonging its lifetime \citep{Meier2023, yoshida_escape_2024}. Additionally, recent work has indicated that atmospheres stripped of their primordial envelopes can pass through a magma ocean phase to eventually yield habitable surface conditions \citep{krissansen_erosion_2024}. The current Hot Rocks survey \citep{diamond_hot_2023}, COMPASS survey \citep{alam_compass_2024, scarsdale_compass_2024} and upcoming STScI Director's Discretionary Time (DDT) Rocky Worlds program will provide a wealth of observations of highly-irradiated rocky exoplanets. Initial results from the Hot Rocks survey do not completely rule out an atmosphere on the exoplanet LHS 1478 b, although the measurements can be explained by a range of degenerate scenarios \citep{august_hot_2024}. JWST observations of L 98-59 d indicate that it may host a high molecular-weight atmosphere abundant in sulphur species \citep{banerjee_atmospheric_2024, gressier_hints_2024}, which may be consistent with volcanic outgassing induced by tidal heating as on the moon Io \citep{seligman_potential_2024, matsuyama_io_2022}. Additionally, there is evidence that the ultra-short-period super-Earth 55 Cancri e hosts a volatile atmosphere composed mostly of \ce{CO2}, which could be be equilibrium with a permanently molten interior \citep{hu_55cnc_2024}. Beyond questions of theory, there is therefore great demand for physical models of the interiors and atmospheres of lava planets, which are necessary for explaining current and upcoming observations.
\par 

In this work we focus on two planets in particular. HD 63433 d is a young Earth-sized exoplanet orbiting a Sun-like star, discovered in transit using TESS \citep{capistrant_hd63433d_2024}. Hydrodynamic models and HST observations indicate that its outer neighbour (HD 63433 b) has already lost its primordial H/He envelope \citep{zhang_escape_2022}. If the more irradiated planet (HD 63433 d) has also lost its envelope then the composition of an overlying secondary atmosphere is likely to be influenced by mantle degassing or volatile exchange with a permanent magma ocean. This hot planet provides an observable analogue for a young Venus \citep{lebrun_thermal_2013, way_venus_2016, turbet_venus_2021} or potentially for Earth shortly following the Moon-forming impact \citep{tonks_magma_1993, canup_moon_2001, warren_moon_1985}.
\par 

TRAPPIST-1 c is an older Earth-sized exoplanet orbiting the M-dwarf star TRAPPIST-1 within a system of seven terrestrial-mass planets \citep{gillon_trappist_2017,agol_trappist_2021}. Secondary eclipse observations by \citet{zieba_no_2023} with JWST/MIRI found a \SI{380 \pm 31}{\kelvin} brightness temperature in the \SI{15}{\micro\meter} \ce{CO2} absorption band, which is potentially consistent with emission by a bare rock surface heated by incoming stellar radiation. This could indicate that TRAPPIST-1 c has no atmosphere, as heat redistribution by atmospheric dynamics are thought to yield a lower brightness temperature. Observations of the planet's inner neighbour by \citet{greene_thermal_2023} more strongly indicate that TRAPPIST-1 b lacks an atmosphere.
\par

\par 
Across the literature, the majority of theoretical studies on lava planet and planets in a runaway greenhouse state have made the assumption that their atmospheres are fully convective. That is: that they are entirely unstable to convection throughout from the surface upwards. This means that the corresponding atmospheric temperature profiles are prescriptively set by dry convection in dry regions, moist convection in regions where volatiles condense, and in some cases through a parametrized isothermal stratosphere \citep{abe_early_1986, ET2008, hamano_lifetime_2015, lebrun_thermal_2013, kopparapu_habitable_2013, Schaefer2016, Salvador2017, zieba_no_2023, krissansen_erosion_2024, boukrouche_beyond_2021, nicholls_proteus_2025}. This assumption is central to our current understanding of the runaway greenhouse transition \citep{graham_multispecies_2021, goldblatt_runaway_2013}. \citet{nicholls_proteus_2025} studied the time-evolved behaviour of young Earth-sized planets at arbitrary redox state, up to the point at which they reach radiative equilibrium -- where a given planet will cease cooling or warming -- or when their initial magma oceans solidify. They employed the coupled interior-atmosphere framework ``PROTEUS'' first developed by \citet{lichtenberg_vertically_2021}. As in previous works,  they made the assumption of complete convective instability by using the ``JANUS'' atmosphere model, which generate atmosphere temperature structures using a moist pseudoadiabat \citep{graham_multispecies_2021}. Their analysis of atmospheric radiative heating rates calculated by JANUS strongly indicated that atmospheres overlying magma oceans may be stable to convection, depending on their composition.

\citet{selsis_cool_2023} used a radiative-convective model of pure steam atmospheres to test the assumption of convective instability, finding that deep isothermal layers -- stable to convection -- can form. The presence of these layers means that these planets may be able to reach radiative equilibrium at lower surface temperatures than previously thought, precluding permanent magma oceans in some cases. \citet{zilinskas_observability_2023} and \citet{piette_rocky_2023} modelled atmospheres containing rock vapours of refractory elements, yielding a range of near-isothermal and inverted atmospheres, which also contrasts with the previously-held assumption of convective stability. 
\par 

% While we do not model escape processes in this work, it is of primary importance to know the potential types of atmospheres that may have been produced by magma ocean solidification, as this will determine the volatile inventory that could be escaped over the planets' lifetime.

In this paper we use numerical simulations explore whether the atmospheres of mixed composition overlying magma oceans remain convectively unstable, and the implications that this has for planetary cooling and magma ocean solidification. We present a new atmosphere model developed for this purpose, which is coupled into an updated version of the PROTEUS framework for magma ocean evolution. We specifically apply this framework to the cases of HD 63433 d and TRAPPIST-1 c, to make assessments about their potential evolutionary histories, and probe the range of evolutionary pathways that Earth-mass planets could exhibit.

\section{Methods}
\label{sec:methods}

\subsection{Interior-atmosphere coupling}
\label{ssec:proteus}
To simulate the evolution of these young planets, we make use of the PROTEUS framework \citep{nicholls_proteus_2025}. PROTEUS couples the interior dynamics code SPIDER \citep{Bower2018,bower_retention_2022} to a choice of atmosphere models, allowing for energy and matter (i.e. volatile gases) to be transported between the interior and atmosphere of a simulated lava planet. This enables us to model the time-evolution of lava planets for a wide range of atmospheric compositions consistent with chemical and solubility equilibrium with the underlying magma ocean. The model terminates either when the mantle solidifies or when the planet reaches a steady state (i.e. global energy balance against irradiation from the host star).
\par

Sulphur geochemistry is thought to be important for the partitioning of elements between the atmosphere and interior on Earth-like planets \citep{lichtenberg_review_2025, kaltenegger_so2_2010}. Additionally, observations of Venus and telescope observations of potentially molten exoplanets have indicated the presence of sulphur species in their atmospheres \citep{taylor_venus_2014, gressier_hints_2024, banerjee_atmospheric_2024}. It has also been suggested that \ce{SO2} could be used as a tracer for exoplanetary volcanism \citep{seligman_potential_2024}.We have therefore updated PROTEUS to include sulphur degassing: the model implements the \ce{S2} solubility law described by \citet{gaillard_redox_2022}, which is also oxidised into \ce{SO2} in the atmosphere with an equilibrium reaction described by \citet{boulliung_so2_2022}. The behaviour of this outgassing model is presented in Section \ref{ssec:outgas}.

\subsection{Time-averaged separation}
\label{ssec:separation}
In \citet{nicholls_proteus_2025} we modelled planets with fixed circular orbits with radii identical to their semi-major axes $a$. However, the eccentricity $e$ of HD 63443 d modelled in this work is relatively large, which modulates the amount of stellar radiation that the planet receives from its host star. To account for this, we use the time-averaged orbital separation
\begin{equation}
    d = a (1 + e^2/2),
    \label{eq:separation}
\end{equation}
to calculate the radiation flux impinging upon the top of its atmosphere.

\subsection{Atmosphere model}
\label{ssec:agni}
AGNI is a new 1D numerical model of the atmospheres of lava planets \citep{nicholls_agni_2025}. Unlike JANUS, it is able to account for the possibility of convective shutdown. AGNI is written in Julia \citep{julialang}, and implements the SOCRATES radiative transfer suite developed by the UK Met Office to simulate the gaseous absorption, collision-induced absorption, and Rayleigh scattering of radiation \citep{edwards_studies_1996, manners_socrates_2017, amundsen_radiation_2014, sergeev_socrates_2023}. This is the same radiative transfer scheme as used in the JANUS atmosphere model. Building upon the work of \citet{nicholls_proteus_2025}, we include the additional sources of opacity outlined in Table \ref{tab:opacity_citations}. The atmosphere model is constructed of $N$ layers (cell-centres), corresponding to $N+1$ interfaces (cell-edges). The radiative transfer calculation takes cell-centre temperatures $T_i$, pressures $p_i$, geometric heights $z_i$, and mixing ratios as input variables at each layer $i$, as well as the surface temperature and incoming stellar flux. Cell-edge quantities in the bulk atmosphere are interpolated from cell-centres using the PCHIP algorithm \citep{fritsch_pchip_1984}. In return from the radiative transfer, we obtain cell-edges spectral fluxes $F_i$ at all $N+1$ interfaces, for upward and downward streams. 
\begin{table}
    \centering
    \begin{tabular}{p{0.08\linewidth}  p{0.17\linewidth} p{0.55\linewidth} }
    \hline
    Gas      & Linelist     & References \\ 
    \hline
    \ce{NH3} & CoYuTe       & \citet{derzi_nh3_2015,coles_nh3_2019}   \\
    \ce{SO2} & ExoAmes      & \citet{underwood_so2_2016}  \\
    \ce{N2O} & HITEMP2019   & \citet{hargreaves_n2o_2019}  \\
    \ce{O3}  & HITRAN2020   & \citet{gordon_hitran_2022}   \\
    \ce{H2S} & AYT2         & \citet{chubb_h2s_2018,azzam_h2s_2016} \\
    \ce{HCN} & Harris       & \citet{harris_hcn_2006, barber_hcn_2013} \\
    \end{tabular}%
    \caption{Additional sources of line-absorption derived from the DACE database \citep{grimm_database_2021}. All other sources of opacity are described in \citep{nicholls_proteus_2025}.}
    \label{tab:opacity_citations}
\end{table}
\par

%{\color{magenta} % revision 1
AGNI also employs a parameterisation of atmospheric convection.
%} 
Convection is a process which occurs in fluids across more than one spatial dimension, although it can play a critical role in energy transport, and so it must be parametrized in our 1D model. We apply mixing length theory (MLT), which allows us to directly calculate the energy flux $F^{\text{cvt}}$ associated with convective heat transport \citep{robinson_temperature_2014, lee_dynamically_2024}.
\begin{equation}
    F^{\text{cvt}} =  \frac{1}{2} \rho c_p w T \frac{\lambda}{H}  (\nabla-\nabla_{\text{ad}}),
    \label{eq:convection}
\end{equation}
where $T$ is the local temperature, $c_p$ is the mass-specific heat capacity of the gas,
\begin{equation}
    w = \lambda  \sqrt{(g/H)  (\nabla-\nabla_{\text{ad}})}
\end{equation}
is the characteristic upward velocity, $H$ is the pressure scale height, 
\begin{equation}
    \nabla = \frac{\di \ln T}{\di \ln P}
\end{equation}
is the local lapse rate, and 
\begin{equation}
    \nabla_{\text{ad}} =  \frac{R}{\mu c_p}
    \label{eq:adiabat}
\end{equation}
is the dry adiabatic lapse rate \citep{pierrehumbert_book_2010}. $\mu$ is the local mean molecular weight of the gas. $R = \SI{8.314463}{\joule \per \kelvin \per \mol}$ is the universal gas constant. This formulation assumes that convecting parcels of gas transport energy by diffusing over a characteristic mixing length
\begin{equation}
    \lambda = \frac{k z} { 1 + k z / H},
\end{equation}
where $k$ is Von Karman's constant and $z$ is the geometrical height from the surface of the planet \citep{hogstrom_karman_1988, blackadar_mlt_1962, joyce_mlt_2023}. In practice, the resultant temperature structures and convective fluxes have little sensitivity to this asymptotic parameterization of $\lambda$ \citep{joyce_mlt_2023}. Using MLT to parameterise convection allows the model to calculate an estimate for the eddy diffusion coefficient of convective mixing,
\begin{equation}
    K_{zz} = w \lambda,
\end{equation}
although this does not account for mixing by other dynamical processes \citep{parmentier_mixing_2013, noti_effects_2024}.
\par

AGNI models the sensible transport $F^{\text{sns}}$ of heat between the surface and the lowest layer of the atmosphere using a simple turbulent kinetic energy scheme
\begin{equation}
    F^{\text{sns}} = c_p \rho C_d U (T_s - T)
    \label{eq:sensible}
\end{equation}
where $C_d=0.001$ is a heat exchange coefficient, and $U=\SI{2}{\meter\per\second}$ is the surface wind speed \citep{pierrehumbert_book_2010}.
\par

Volatile phase change (condensation, sublimation, evaporation) and associated energy transport is included through a scheme which assumes a fixed condensation timescale $\tau$. Taking $\tau=0$ is physically equivalent to instantaneous moist adjustment as in \citet{selsis_cool_2023, innes_runaway_2023}. Given a temperature profile, this takes place as follows:
\begin{enumerate}
    \item A temperature profile is calculated or provided (such as by equation (\ref{eq:temperatures})).
    
    \item Based on the atmosphere temperature-pressure points, condensation of super-saturated condensable volatiles occurs at each pressure level $i$ until the partial pressures $p_j$ of all volatiles $j$ are equal to their saturation pressure $p_j^{\text{sat}}(T_i)$ at the local temperature $T_i$. The corresponding change in gas volume mixing ratio is denoted $\delta \chi_j$.
    
    \item The mixing ratios of dry (non-condensing) species are increased in order to satisfy the total pressure at condensing layers. This is consistent with the model using a fixed pressure-grid \citep{innes_runaway_2023}.
    
    \item For volatiles $j$ which were super-saturated according to the temperature profile in the first step and then set to saturation in the second step, a corresponding amount of condensate [\SI{}{\kilo\gram\per\meter\squared}] is generated at $i$ by mass conservation:
    \begin{equation}
        \delta m_{j,i} = \mu_j p_i \frac{\delta \chi_j}{ g_i \mu_i}.
    \end{equation}
    
    \item The mass of condensate $\delta m_{j,i}$ is calculated at each layer $i$. The sum of this quantity for a given $j$ represents the total amount of condensate (rain) produced throughout the column. The sum of $\delta m_{j,i}$ over all condensing regions is distributed into dry regions at pressure levels below, under the assumption that no clouds are formed. $\delta m_{j,i}$ is assigned a negative sign in these evaporating regions, such that $\sum_i \delta m_{j,i} = 0$ for a given volatile $j$.
    
    \item The local heating associated with phase change in condensing and evaporating regions is then calculated as
    \begin{equation}
        dF^{\text{lat}}_{j,i} = \delta m_{j,i} L_{j}(T_i) / \tau.
    \end{equation}
    
    \item $dF^{\text{lat}}$ is then integrated from the top of the atmosphere downwards to provide cell-edge latent heat transport fluxes $F^{\text{lat}}$ between adjacent layers. 
\end{enumerate}
The latent heat flux is always positive, and has a value of zero at the top and bottom of the model. Since condensation is taken to be part of a cycle of upward transport, condensation, precipitation, and evaporation, this latent heat transport is effectively generated by the motion of condensable gases \citep{pierrehumbert_book_2010, ding_convection_2016}. The timescale $\tau$ physically represents the microphysics of condensation. Calculating $\tau$ self-consistently with a microphysics model is beyond the scope of this work, so we use a fixed value of \SI{3e4}{\second} \citep{trenberth1992climate}. This scheme for handling condensation is preferred over hard moist adjustment, as it is does not requiring invocation of enthalpy conservation constraints, is differentiable with respect to temperature -- a necessary requirement for the nonlinear solution method introduced below -- and could be extended in future work to include a microphysics-informed calculation of $\tau$. Heat capacities $c_p(T)$ are temperature dependent with values derived from the JANAF tables. The saturation pressures $p^{\text{sat}}(T)$ and enthalpies of phase change $L(T)$ for all condensable gases are also temperature-dependent \citep{IAPWS95, feistel_ice_2006, coker_thermo_2007}. The atmosphere is assumed to be a hydrostatically supported ideal gas. 
\par

The energy flux $F_i = F_i^{\text{up}} - F_i^{\text{down}}$ describes the net upward-directed energy transport [\SI{}{\WPMS}] from layer $i$ into the adjacent layer $i-1$ above (or into Space for $i=1$). Taking all of the energy transport processes together at $i$, this is simply:
\begin{equation}
    F = F^{\text{sns}} + F^{\text{cvt}} + F^{\text{lat}} +  F^{\text{rad,up}}  -  F^{\text{rad, down}} 
\end{equation}
For energy to be conserved throughout the column, under the plane parallel approximation it must be true that $F_i = F_t \text{ } \forall \text{ } i$ where $F_t$ is some unknown constant to be determined. We use this construction in AGNI to calculate the temperature profile of the atmosphere as an $N+1$-dimensional root-finding problem. Solving for the $T(p)$ profile which conserves energy fluxes in this way avoids having to invoke heating rate calculations, thereby sidestepping slow convergence in regions with long radiative timescales. The temperature structure solution is obtained iteratively from some initial guess. The residuals vector (length $N+1$) of flux differences
\begin{equation}
    \bm{r} = 
    \begin{pmatrix}
            r_i     \\
            r_{i+1} \\
            \vdots     \\
            r_N     \\ 
            r_{N+1} 
    \end{pmatrix} = 
    \begin{pmatrix}
        F_{i+1} - F_i     \\
        F_{i+2} - F_{i+1} \\
        \vdots     \\
        F_{N+1} - F_N     \\ 
        F_{N+1} - F_t
    \end{pmatrix}
    \label{eq:residuals}
\end{equation}
is calculated using the temperature guess,
\begin{equation}
    \bm{x} = 
    \begin{pmatrix}
        T_i     \\
        T_{i+1} \\
        \vdots     \\
        T_N     \\ 
        T_s
    \end{pmatrix}
    \label{eq:temperatures}
\end{equation}
where $T_s$ is the surface temperature. The bottom- and top-most cell edge temperatures are extrapolated log-linearly with $dT/d \log p$, although these layers are set to be very small to avoid numerical artifacts.
\par

The Jacobian matrix $\bm{J}$ represents the directional gradient of the residuals $\bm{r}$ with respect to the solution vector $\bm{x}$. It is a square matrix with elements defined as
\begin{equation}
    J_{uv} = \frac{\partial r_u}{\partial x_v}   
    \label{eq:jacobian}
\end{equation}
AGNI estimates $\bm{J}$ using finite-differences: each level $v$ with temperature $x_v$ is perturbed by an amount $\pm \varepsilon x_v$ in order to fill a single column of $\bm{J}$. It can be expensive to construct a full Jacobian and would be wasteful if it were discarded at the end of each iteration, so AGNI retains some of the columns in $\bm{J}$ between iterations. This assumes that the second derivative of the residuals is small. A column $v$ is retained only when 
\begin{equation}
    \max_i |r_i| < 0.1 \text{ for } i \in \{v-2, v, v+2\} 
\end{equation}
and when the atmosphere is not near convergence. With a Jacobian constructed, an update $\bm{d}$ to the solution vector $\bm{x} \rightarrow \bm{x} + \bm{d}$ is performed. This is calculated using the Newton-Raphson method $\bm{d} = -\bm{J}^{-1} \bm{r}$. If the model enters a local minimum, it is nudged by scaling the update $\bm{d} \rightarrow s \bm{d}$ by a factor $s$. The non-convexity of the solution space and somewhat discontinuous nature of the physics means that large $\bm{d}$ can be problematic. When $|\bm{d}| > d_{\text{max}}$, the update is scaled as $\bm{d} \rightarrow  d_{\text{max}} \hat{\bm{d}}$. To improve convergence, the update may also be scaled by a linesearch method $\bm{d} \rightarrow \alpha \bm{d}$. This is applied if the full step $\bm{d}$ would increase the cost by an unacceptable amount. If the model is close to convergence then a golden-section search method is used to determine $\alpha$, otherwise a backtracking method is used. All three of these scalings to $\bm{d}$ preserve its direction through the solution space. The model converges when the `cost function',
\begin{equation}
    \mathcal{C}(\bm{x}) = \sqrt{\sum_i |r_i|}
\end{equation}
satisfies the condition 
\begin{equation}
    \mathcal{C}(\bm{x}) < \mathcal{C}_a + \mathcal{C}_r \max_i \text{ } |F_i|
\end{equation}
which represents a state where the fluxes are sufficiently conserved. $\mathcal{C}_a \text{ and } \mathcal{C}_r$ are the absolute and relative solver tolerances. 
\par 

In this work, atmospheres are modelled with well-mixed gas compositions (except for the potential rainout of volatiles during with condensation). It is possible to couple AGNI self-consistently to the equilibrium chemistry code ``FastChem'' \citep{stock_faschem_2022, stock_fastchem_2018}. This is done by assuming that the atmosphere is elementally well-mixed in order to calculate its metallicity. Then, for each estimate of the temperature profile $\bm{x}$, FastChem is used to calculate the speciation of the elements (H, C, N, O, S) at each level of the atmosphere. The additional species modelled by FastChem contribute to the atmospheric mean molecular weight, heat capacity, and opacity, where data are available. Energy transport fluxes are calculated in the same manner as described above for atmospheres with isochemical compositions; this allows for feedback between the chemistry and the energy transport processes \citep{drummond_effects_2016, nicholls_temperaturechemistry_2023}. Since FastChem is a model of equilibrium chemistry, this does not account for the diffusive mixing of material, photochemistry, or the generation of excess enthalpy by disequilibrium processes. Coupling between AGNI and FastChem is only applied within this work in Section \ref{ssec:chem}.
\par 

Our evolutionary calculations assume the presence of a thin conductive boundary layer at the surface of the magma ocean \citep{solomatov_treatise_2007,Schaefer2016,Bower2018,lichtenberg_vertically_2021, nicholls_proteus_2025}. The layer conducts a heat flux according to Fourier's law
\begin{equation}
    F_t = \kappa_c \frac{T_m-T_s}{d_c}
    \label{eq:conduction}
\end{equation}
where $T_m$ is the magma temperature immediately below the boundary layer, providing the boundary condition for the bottom of the atmosphere.
\par 

This new atmosphere model is validated against previous work in Section \ref{ssec:agni}.

\subsection{Planets}
\label{ssec:planets}
We use PROTEUS to simulate the time evolution of two rocky exoplanets: HD 63433 d and TRAPPIST-1 c. Table \ref{tab:planets} outlines the relevant physical parameters for these planets and their host stars.
\begin{table}
\centering
\begin{tabular}{p{0.3\linewidth} p{0.25\linewidth} p{0.25\linewidth} }
\hline 
Planet                   & HD 63433 d       & TRAPPIST-1 c \\
\hline
Mass [$M_\oplus$]        & 1.0$^\dag$       & 1.308    \\
Radius [$R_\oplus$]      & 1.073            & 1.097       \\
Semi-major axis [AU]     & 0.503            & 0.0158     \\
Eccentricity             & 0.16             & 0.00654       \\
Equilibrium temp. [K]    & 1040             & 313 \\
\hline
Star                     & HD 63433         & TRAPPIST-1   \\ 
\hline
Class                    & G5V              & M8V        \\
Mass [$M_\odot$]         & 0.99             & 0.10$^\dag$ \\
Current age [Myr]        & 414              & 7600       \\
Spectral analogue        & Sun$^\dag$       & TRAPPIST-1  \\ 
\end{tabular}%
    \caption{Planetary and stellar parameters used in this study \citep{capistrant_hd63433d_2024, grimm_trappist_2018, gillon_trappist_2017, agol_trappist_2021}. Daggers$^\dag$ indicate values for which we have made necessary physical assumptions. The mass of HD 63433 d is unknown, so we adopt a value of $1.0 M_\oplus$ for simplicity. It is unlikely that this mass is an underestimate because there are currently no low-density (i.e. sub-Earth) exoplanets with Earth-like radii \citep{parc_from_2024}. We define mass and radius at the atmosphere-interior boundary. We adopt the solar (G2V) spectrum as an analogue for HD 63433 \citep{gueymard_suns_2004}, and the TRAPPIST-1 (M8V) Mega-MUSCLES spectrum for TRAPPIST-1 \citep{froning_muscles_2019, wilson_muscles_2021}. TRAPPIST-1 is modelled with a mass of $0.1 M_\odot$ -- rather than $0.0898 M_\odot$ -- as $0.1 M_\odot$ is the lower-limit on the stellar masses covered by the \citet{spada_radius_2013} evolution tracks used in this work. Equilibrium temperatures taken from the literature are scaled such that they both correspond to a Bond albedo of 30 per cent.}
    \label{tab:planets}
\end{table}
We assume nitrogen (2.01 ppmw) and sulphur (235.00 ppmw) abundances and a bulk C/H mass ratio consistent with Earth's primitive mantle \citep{wang_elements_2018}. The radius of the metallic core is fixed at 55 percent of the radius of the atmosphere-mantle interface \citep{lodders_planetary_1998}. The total hydrogen inventory is set equivalent to eight Earth oceans \citep{salvador_venus_2023, lichtenberg_review_2025}. The surface of the planet has a spectrally grey albedo of 30 per cent \citep{Essack_2020, Fortin_2024}. As per \citet{nicholls_proteus_2025}, the stellar spectra are evolved self-consistently with the time-evolution of the planets using the evolution tracks of \citet{spada_radius_2013} and the spectral model of \citet{johnstone_active_2021}. Both systems are initialised at a stellar age of \SI{100}{\mega\year}, which results in enhanced instellations compared to present-day values \citep{baraffe_new_2015, spada_radius_2013}. Both planets are likely to be tidally-locked and synchronously rotating, so to represent the global behaviour of the planet with a single column we use a zenith angle of $\cos^{-1}(1/\sqrt3) = 54.74^{\circ}$ and scale the stellar spectrum by a factor of $1/4$ as per \citet{hamano_lifetime_2015}. 
\par

For each planet, we vary the mantle oxygen fugacity ($f\ce{O2}$) between -5 and +5 log units relative to the Iron-W\"ustite (IW) buffer. $f\ce{O2}$ is held constant throughout each simulation. In total, 44 simulations are run, corresponding to all combinations of: planet (HD 63433 d, TRAPPIST-1 c), atmosphere model (AGNI, JANUS), and $f\ce{O2}$ (11 samples). This allows us to explore the impact of $f\ce{O2}$ on early planetary evolution, and to explore whether characterisation of the atmospheres of lava worlds can offer insight into their interiors. These simulations are presented in Sections \ref{ssec:evol}-\ref{ssec:emit}. These simulations are conducted without coupling to FastChem, except in Section \ref{ssec:chem}.

\section{Results}
\label{sec:results}

\subsection{Outgassing model}
\label{ssec:outgas}
\begin{figure}
    \centering
    \includegraphics[width=\linewidth, keepaspectratio]{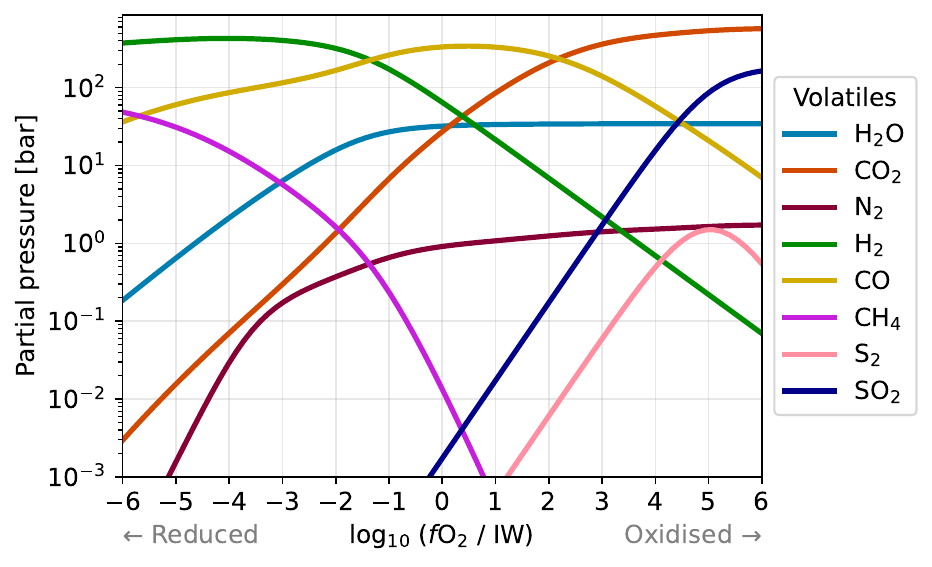}%
    \caption{Gas partial pressures at \SI{2500}{\kelvin} versus $f\ce{O2}$. Mid-ocean ridge basalt has an $f\ce{O2}$ approximately equal to the Fayalite-Magnetite-Quartz (FMQ) buffer, equivalent to IW+3.69 \citep{schaefer_review_2018}. Observations indicate that Mercury's surface probably erupted with an $f\ce{O2}$ between IW-6.5 and IW-3.5 \citep{namur_mercury_2016}, overlapping with the $f\ce{O2}$ of the Solar nebula at approximately IW-7 to IW-6 \citep{Doyle2019,Grewal2024}.}
    \label{fig:outgas}
\end{figure}

To demonstrate the behaviour of sulphur outgassing in the  context of all the other modelled volatiles, we first calculate the partial pressures of the eight volatiles in equilibrium with a magma ocean of fixed mass and temperature, but at variable $f\ce{O2}$. This is plotted in Fig. \ref{fig:outgas}. These partial pressures correspond to potential secondary atmospheres on HD 63433 d in equilibrium with an underlying global magma ocean with 100 per cent melt fraction, subject to equilibrium chemistry in the atmosphere and volatile dissolution into the melt. \ce{CO2} dominates at oxidising conditions ($f\ce{O2}>$IW+2), with \ce{SO2} becoming the second most abundant gas in the most oxidising cases ($f\ce{O2}\gtrsim$4.5). Near IW+0, \ce{CO} becomes the most abundant gas and its volume mixing ratio peaks. Under reducing conditions ($\lesssim$IW-1.5), \ce{H2} dominates the composition and the partial pressure of \ce{CH4} increases significantly for decreasing $f\ce{O2}$. These behaviours and variety of compositions are consistent with \citet{gaillard_redox_2022}, \citet{Suer2023}, and \citet{gillmann_controls_2024}. 

\subsection{Atmosphere model validation}
\label{ssec:valid}
\begin{figure}
    \centering
    \includegraphics[width=\linewidth, keepaspectratio]{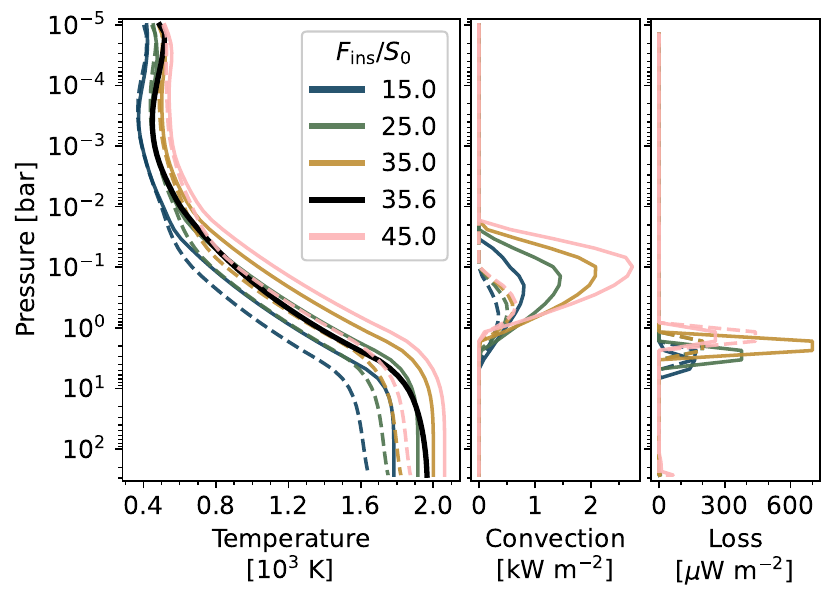}%
    \caption{Models of an Earth-sized planet with a \SI{270}{\bar} pure-steam atmosphere. The \citet{gueymard_suns_2004} solar spectrum is used. Coloured lines plot AGNI atmosphere models calculated using opacities derived from POKAZATEL/DACE (solid lines) and HITRAN (dashed lines), for a range of instellation fluxes (legend). The black line shows the \citet{selsis_cool_2023} model. Left panel: temperature, $T$ in equation (\ref{eq:temperatures}). Centre panel: convective energy flux, $F^{\text{cvt}}$ in equation (\ref{eq:convection}). Right panel: energy lost across each level, $r$ in equation (\ref{eq:residuals}).}
    \label{fig:valid}
\end{figure}

To test AGNI's ability to resolve convectively stable regions and conserve energy, we first apply it standalone to pure-steam atmospheres similar to \citet{selsis_cool_2023}. We consider an Earth-sized planet orbiting the Sun, with an atmosphere of \SI{270}{\bar} of \ce{H2O} (corresponding to a water inventory of approximately one Earth ocean). AGNI is used to solve for radiative-convective equilibrium with $F_t=0$ for a range of instellation fluxes $F_{\text{star}}$ between 15 and 45 times Earth's instellation $S_0 = \SI{1360.8}{\WPMS}$ \citep{kopp_solar_2011}. The opacities are also varied to compare results using the POKAZATEL/DACE database with HITRAN. Fig. \ref{fig:valid} plots the model results: the left panel shows the temperature profiles calculated by AGNI at various instellation fluxes (coloured lines; solid with POKAZATEL and dashed with HITRAN) compared to that from \citet{selsis_cool_2023}. It is immediately clear from this panel that deep isothermal layers form at pressures $p \gtrsim \SI{4}{\bar}$. The temperature profiles are qualitatively similar to those calculated by \citet{selsis_cool_2023}, particularly in the upper atmosphere. AGNI model finds a surface temperature $\SI{35}{\kelvin}$ hotter than theirs for a comparable $F_{\text{ins}} = 35 S_0$. We take this to be reasonable, given the significant differences between the models: convection parameterization (mixing length theory instead of dry adjustment), water opacity (POKAZATEL instead of HITEMP), spectral resolution (256 instead of 69 bands), heat capacity (temperature dependent instead of fixed). The left panel of Fig. \ref{fig:valid} also shows that calculations with POKAZATEL and HITRAN models show minimal differences at low pressures and temperatures -- where HITRAN is sufficiently complete -- but deviate as expected under less temperate conditions near the surface of the planet and in the deeper atmosphere. 
\par 

The middle panel of Fig. \ref{fig:valid} plots convective flux (defined in equation (\ref{eq:convection})) versus pressure, for the different instellations considered. Convection is driven by absorption of shortwave radiation, so it scales with $F_{\text{ins}}$ and with the increased shortwave opacity introduced by POKAZATEL over HITRAN. The right panel of the figure plots flux losses (residuals; equation (\ref{eq:residuals})) at each level calculated by AGNI. This shows that per-level flux losses are small in all cases, up to the numerical tolerance set for the calculation.

\subsection{Evolutionary outcomes}
\label{ssec:evol}

% These two 'figures' are included 
%   within the same {figure} in order 
%   to make sure that they stay together 
%   on the page, and do not drift.
% See the StackExchange post here:
%   https://tex.stackexchange.com/a/134172 
\begin{figure} 
    \centering
    \includegraphics[width=\linewidth, keepaspectratio]{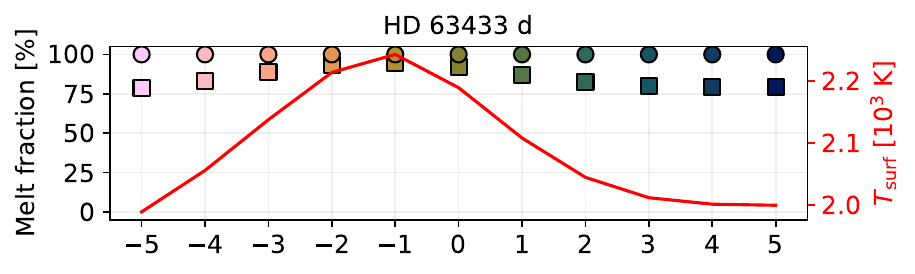}
    \\
    \vspace{-2mm}
    \includegraphics[width=\linewidth, keepaspectratio]{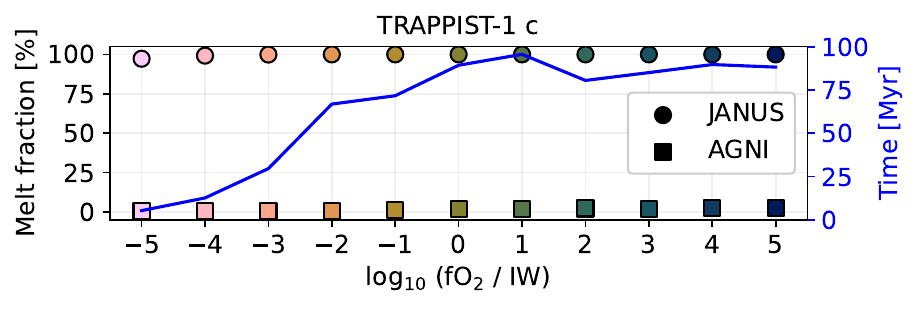}%
    \caption{Mantle melt fraction $\Phi$ at model termination for HD 63433 d (top) and TRAPPIST-1 c (bottom), versus mantle $f\ce{O2}$, simulated with both atmosphere models (marker shape). 
    %{\color{magenta} % revision 1
    The red line on the top plot shows the final surface temperature for the (AGNI) HD 63433 d cases, which all rapidly reach radiative equilibrium. The blue line on the bottom plot shows the time taken for the (AGNI) TRAPPIST-1 c cases to solidify.}
    %}
    \label{fig:melt}
    
    \vspace*{\floatsep}
    
    \includegraphics[width=\linewidth, keepaspectratio]{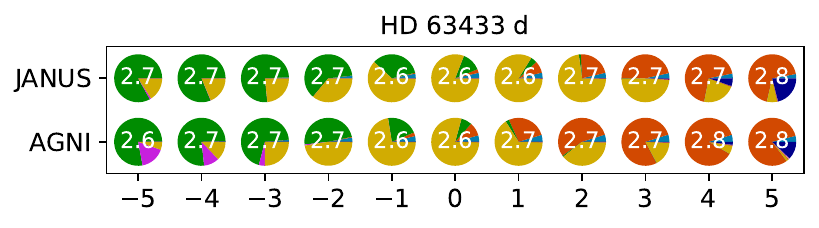}
    \\
    \vspace{-2mm}
    \includegraphics[width=\linewidth, keepaspectratio]{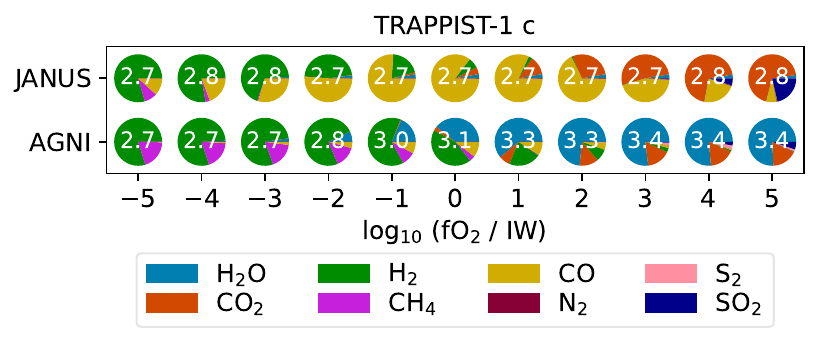}%
    \caption{Outgassed atmospheric composition at model termination for HD 63433 d (top) and TRAPPIST-1 c (bottom), versus mantle $f\ce{O2}$ (x-axis), simulated with both atmosphere models (y-axis). Pie chars show volatile volume mixing ratios and white numbers show $\log_{10}$ total surface pressure [bar].}
    \label{fig:pies}
\end{figure}

Fig. \ref{fig:melt} shows how the melt fractions (y-axes) calculated in our simulations depend on the oxygen fugacity of the mantle (x-axes) and the atmosphere model (top and bottom panels). The top panel shows this for HD 63433 d, where square scatter points show the resultant melt fraction from simulations coupled to AGNI, while circular scatter point show the results from those coupled to JANUS. The bottom panel shows the corresponding results from the TRAPPIST-1 c simulations.
\par 

The top panel of Fig. \ref{fig:melt} shows that simulations of HD 63433 d with the JANUS atmosphere module yield an almost fully molten mantle regardless of the mantle's redox state (x-axis) and corresponding atmospheric composition. Simulations of this planet with the AGNI atmosphere module result in relatively smaller melt fraction (i.e. the mantles have begun to solidify). However, all of the AGNI cases still maintain a significant amount of melt, which is shown to be maximised at IW-1. A permanent magma ocean is therefore possible on HD 63433 d, in spite of its \SI{1040}{\kelvin} sub-solidus planetary equilibrium temperature (Table \ref{tab:planets}), due to insulation provided by the overlying atmosphere. The upper panel of Fig. \ref{fig:pies} plots the corresponding atmospheric compositions, with $f\ce{O2}$ varying on the x-axis and the pie charts indicating atmospheric composition and $\log_{10}$ surface pressure. Models with AGNI and JANUS both vary between \ce{H2} and \ce{CO2} dominated, with little atmospheric \ce{H2O} due to its favourable dissolution into the large amount of melt \citep{nicholls_proteus_2025}.
\par 

There is variability of $\sim \SI{250}{\kelvin}$ in the surface temperature of the magma ocean on HD 63433 d. This is plotted as black line in the top panel of Fig. \ref{fig:melt}. As with melt fraction, surface temperature is maximised near IW-1.
\par 

In the case of TRAPPIST-1 c, simulations with AGNI all solidify: shown by the square scatter points in the bottom panel of Fig. \ref{fig:melt}. This is in complete contrast to simulations with JANUS, which yield permanent magma oceans with large melt fractions (circular points). Comparing with the compositions plotted in the bottom panel of Fig. \ref{fig:pies}, solidification under reducing conditions (small $f\ce{O2}$) does not yield a steam-dominated atmosphere; degassed hydrogen is primarily speciated into \ce{H2} and \ce{CH4}. It is only when the planet solidifies under oxidising conditions that a large \ce{H2O} atmosphere is outgassed, which means that it cannot be a general outcome of magma ocean solidification.  Across all AGNI models, the largest surface pressure (\SI{2661.1}{\bar}) is generated by solidification of TRAPPIST-1 c at IW+5 (bottom panel of Fig. \ref{fig:pies}). The smallest surface pressure (\SI{399.7}{\bar}) corresponds to the permanent magma ocean on HD 63433 d at IW-5. 
\par 

The black line in the bottom panel of Fig. \ref{fig:melt} plots solidification time versus $f\ce{O2}$ for the AGNI cases (which all solidify). This shows that the magma ocean on TRAPPIST-1 c solidifies across a range of timescales from 5.3 to \SI{95.7}{\mega\year} depending on the mantle redox state. Solidification time generally increases under more oxidising conditions where the atmosphere becomes \ce{H2O}-dominated and the greenhouse effect is stronger. This range is too short for significant stellar evolution to occur, and is comparable with that found by \citet{nicholls_proteus_2025}.
\par

\subsection{Temperature and convection}
\label{ssec:prof}
\begin{figure}
    \centering
    \includegraphics[width=\linewidth, keepaspectratio]{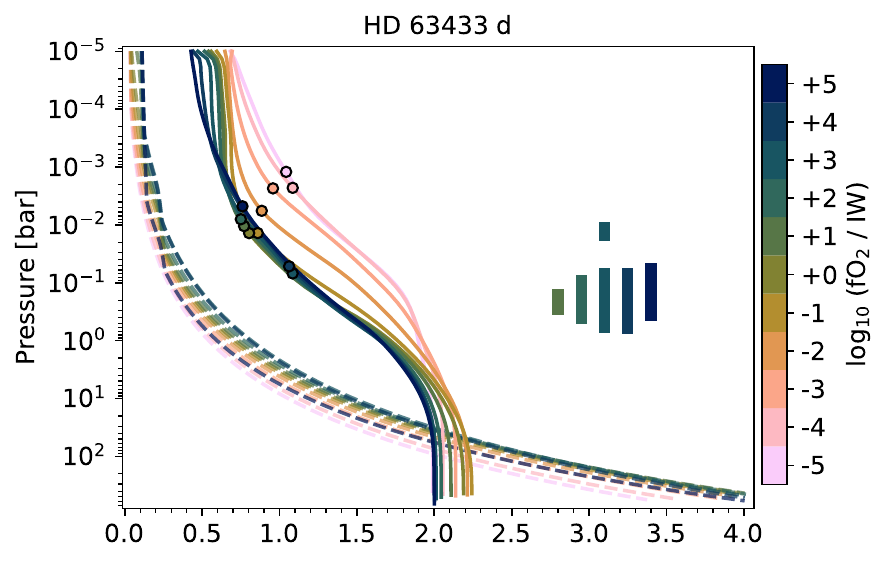}
    \\
    \vspace{-2mm}
    \includegraphics[width=\linewidth, keepaspectratio]{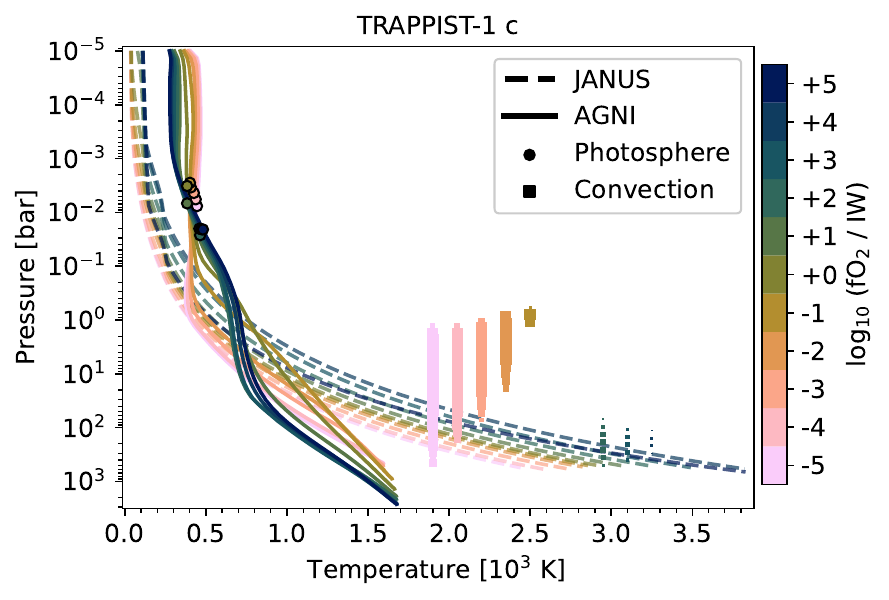}%
    \caption{Atmospheric temperature profiles at model termination for HD 63433 d (top) and TRAPPIST-1 c (bottom), versus mantle $f\ce{O2}$ (line colour), simulated with both atmosphere models (line style). 
    %{\color{magenta} % revision 1
    Square markers denote the presence of convective regions in the AGNI cases, where the marker sizes scale with the relative amount of the total flux at each pressure level that is carried by convection; the horizontal positioning of these markers is for visualisation purposes. At the full square marker size (legend entry), convection is responsible for all upward flux transport at that level. Circular markers denote the effective photospheres, taken to be the pressure levels at which the contribution functions are maximised in each case \citep{knutson_multi_2009, drummond_effects_2016, boukrouche_beyond_2021}. Note the different axis limits between the subplots.}
    %}
    \label{fig:prof}
\end{figure}

Let us now consider the behaviour of the atmosphere above the surface. Fig. \ref{fig:prof} plots atmospheric temperature-pressure profiles for all cases considered: the top panel for HD 63433 d and the bottom for TRAPPIST-1 c. As with the melt fraction, these strongly depend on the mantle redox ($f\ce{O2}$, colourbar) as it exerts significant control over the composition of the atmospheres. Profiles produced by AGNI are shown with solid lines, while the pseudoadiabats produced by JANUS are shown with dashed lines. The square scatter points indicate regions in which convection is occurring in the atmosphere, with the same colour mapping. Note that these all correspond to atmospheres in equilibrium with permanent magma oceans, with the exception of TRAPPIST-1 c when modelled with AGNI (Section \ref{ssec:evol}).
\par 

For HD 63433 d, the top panel of Fig. \ref{fig:prof} shows that convection only persists at radiative-equilibrium at pressures $p < \SI{1}{\bar}$ and mantle $f\ce{O2} \ge \text{IW+1}$. These regions of convection are enabled by the absorption of downwelling stellar radiation, which is back-scattered more strongly for the reducing cases that have higher amounts of \ce{H2} and \ce{CO}. Convection is the dominant process transporting energy upwards through these regions of the atmosphere (scatter points in top panel), while energy transport in deeper layers is handled by radiation alone. Indeed, under reducing conditions the atmospheres are entirely radiative. This shows that convection can shut down even for atmospheres of mixed composition. Despite this convective shutdown, the surface temperatures on HD 63433 d are still sufficiently high to support a permanent magma ocean, even with a sub-solidus equilibrium temperature (Table \ref{tab:planets}). This is possible because the radiative-convective AGNI models have shallower temperature profiles (in $\di \ln(T)/ \di \ln(P)$ terms) than the pseudoadiabatic JANUS models, which allows for smaller temperature contrasts between the planet's surface and the radiating layer aloft, which means that they can still attain radiative equilibrium above a permanent magma ocean. Grey analytic solutions \citep{guillot_2010, pierrehumbert_book_2010} have shown that the temperature of deep isothermal layers depends on the pressure-thickness of the region between the effective shortwave-absorbing layer and the thermal photosphere; when this region is thick, the deep isothermal layer can become very hot. The results presented in Fig. \ref{fig:prof} -- using spectrally-resolved radiative transfer calculations -- are therefore in line with expectations from grey analytic solutions. The temperature of the upper-atmospheres modelled by AGNI are found to increase under more reducing conditions. Condensation of volatiles and strong temperature inversions are absent throughout these simulations.

\par 
The square scatter points in the bottom panel of Fig. \ref{fig:prof} shows that deep convection persists in the models of a young TRAPPIST-1 c under reducing ($\le \text{IW}-1$) conditions. These profiles do not sit exactly on the dry adiabats (dashed lines) within convective regions because radiative fluxes are non-zero, and therefore still influence the temperature structures within these regions. It is important to note that the AGNI models (solid lines) of this planet correspond to cases with positive $F_t$ (see equation (\ref{eq:conduction})) as the planet has not reached radiative equilibrium and would continue to cool post-solidification. As with HD 63433 d, the upper-atmosphere profiles also shows a similar trend of increasing temperature under more reducing conditions. Condensation and temperature inversions are both absent, although the reducing cases yield a tall near-isothermal stratosphere which corresponds to a relatively Planckian emission spectrum (bottom panel of Fig. \ref{fig:emit}).

\subsection{Emission spectra}
\label{ssec:emit}
\begin{figure*}
    \centering
    \includegraphics[width=0.93\linewidth, keepaspectratio]{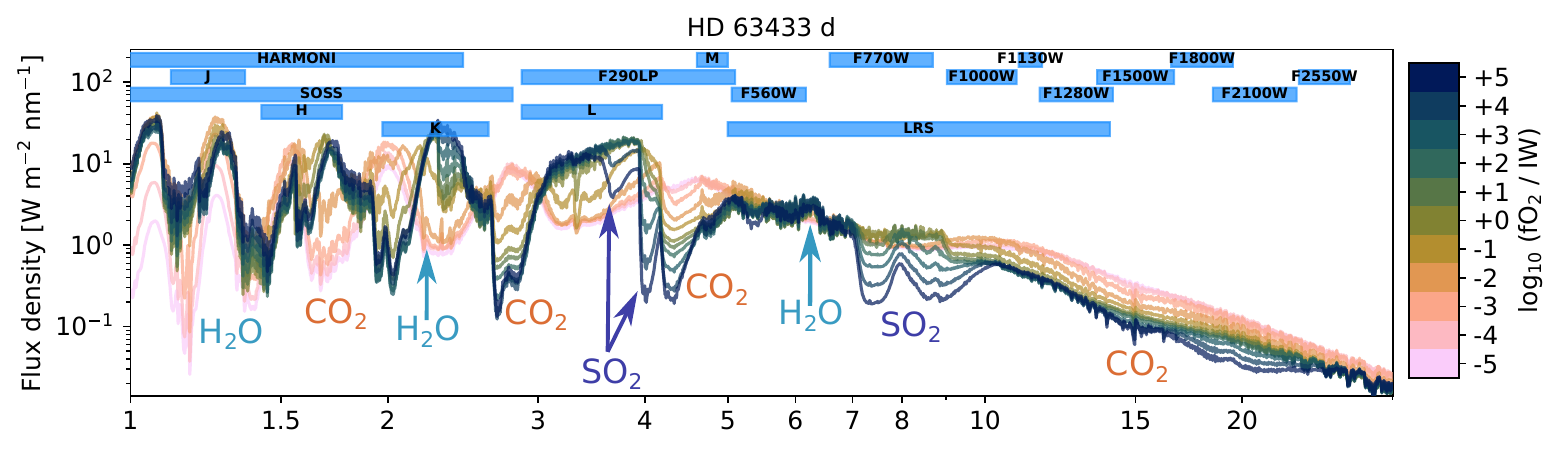}
    \\
    \vspace{-2mm}
    \includegraphics[width=0.93\linewidth, keepaspectratio]{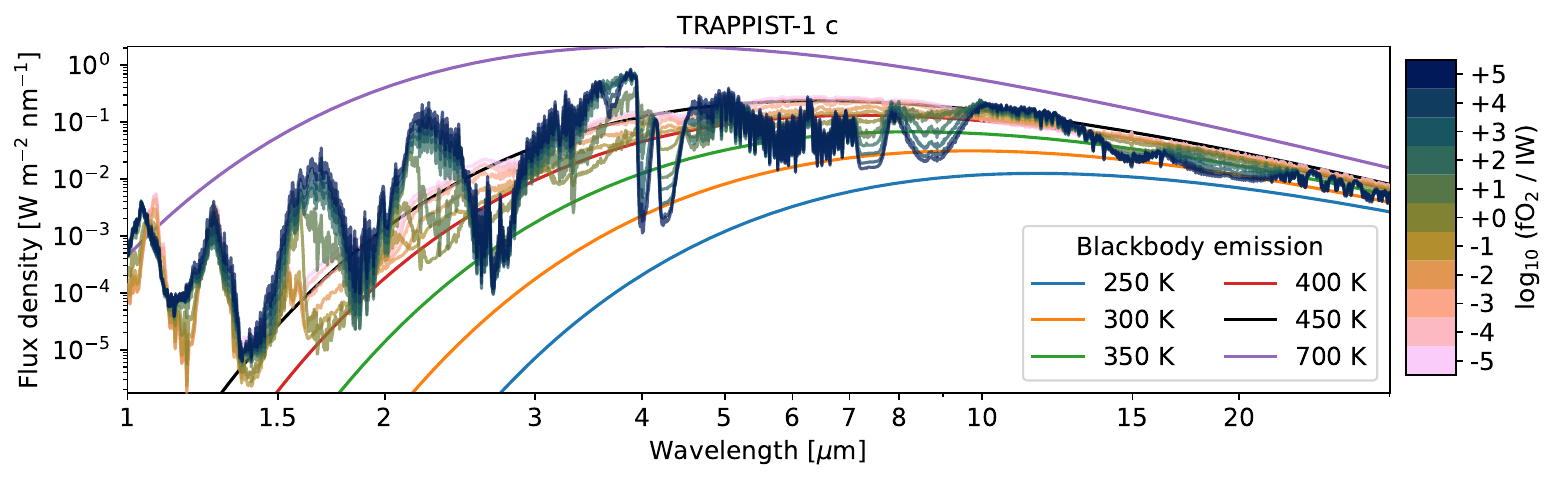}%
    \caption{Top-of-atmosphere flux representative of the planet-averaged outgoing radiation at model termination for HD 63433 d and a young TRAPPIST-1 c, versus mantle $f\ce{O2}$ (colourbar). For HD 63433 d (top panel) this corresponds to the planet at radiative equilibrium at an early stage of its life. For TRAPPIST-1 c (bottom panel) this corresponds to the planet at the point of magma ocean solidification. Note that these do not represent models of the present-day emission spectra; they not directly comparable with specific telescope observations. Radiative transfer is calculated with AGNI/SOCRATES at an increased spectral resolution (4096 bands) using the same opacities as in the evolutionary calculations. Outgoing radiation includes thermal emission from the atmosphere and surface, back-scattered stellar radiation, and surface reflection. Blue rectangles in the top panel indicate some of the bandpasses for MIRI, NIRSpec, NIRISS, HARMONI, and common high-resolution bands. Legend lines in the bottom panel plot the blackbody emission at various brightness temperatures.}
    \label{fig:emit}
\end{figure*}

The cooling rate of these model planets is determined by radiative transport of energy to space. This inherently provides synthetic emission spectra which include the combined contributions of atmospheric thermal emission, back-scattered stellar radiation, and surface reflection. Fig. \ref{fig:emit} plots top of atmosphere infrared emission spectra for both planets (top and bottom panels) simulated with AGNI, for each of the eleven oxidation states considered (colour bar). It is important to note that, because we use a 1D column model, these emission spectra represent the outgoing radiation from the planet as a whole, and do not the dayside emission that would be observed from a secondary eclipse. Dayside (or even substellar) emission could be more sensitive to the redistribution of heat due to zonal dynamics within the atmosphere, and therefore would require specific modelling that is beyond the scope of this work \citep{hammond_rotational_2021, koll_scaling_2022}.
%{\color{magenta} % revision 1
The corresponding emission spectra arising from simulations with JANUS are presented and discussed in Appendix \ref{app:emit_janus}.
%}
\par 

These spectra all show several molecular features; notably from \ce{CO2}, \ce{SO2}, and \ce{H2O}, which are annotated on the top panel of the plot. These features are all generated by upper-atmospheric absorption of radiation produced by the deeper -- hotter -- atmosphere, as the temperature profiles are uninverted (c.f. Fig. \ref{fig:prof}) and we assume a spectrally grey surface emissivity. For both planets (top and bottom panels), the presence and depth of the features varies with $f\ce{O2}$ (colour bar). A smaller upwelling flux within the \ce{CO2} and \ce{SO2} features in Fig. \ref{fig:emit} correlates with an oxidised mantle due to the difference in the composition of the outgassed atmosphere. Oxidising mantles have been shown to yield atmospheres rich in \ce{CO2} and \ce{SO2} (see Fig. \ref{fig:pies}), and thus have absorption features associated with these gases. Molecular features are less discernible at wavelengths greater than \SI{20}{\micro\meter}, although outgoing flux generally increases with upper-atmosphere temperature (Section \ref{ssec:prof}). 

\par 
The absorption features fall within the bandpasses of JWST's instruments: MIRI (LRS and photometry), NIRSpec (F290LP). The \ce{SO2} feature at \mbox{7-\SI{9}{\micro\meter}} corresponds closely with MIRI's F770W filter, and is strongly associated with deviation from the Planck function under the oxidising conditions in which sulphur outgassing is favoured. There are also features at shorter wavelengths within the bandpass for direct imaging with the upcoming ELT. 

\par 
The bottom panel shows the synthetic emission spectra for a young TRAPPIST-1 c. Caution should be taken when comparing these to recent observations of this planet \citep{zieba_no_2023}, as Fig. \ref{fig:emit} does not plot dayside emission spectra and does not account for physical processes that may have occurred on TRAPPIST-1 c between the solidification of its magma ocean and the present day. However, there are there are clear molecular features associated with mantle redox, as with HD 63443 d. Within the \ce{CO2} feature at \SI{15}{\micro\meter}, the brightness temperature varies between $\sim 300$ and \SI{450}{\kelvin}. Which brackets the \SI{380}{\kelvin} dayside value detected by \citet{zieba_no_2023}.

\par 
There are no emission features directly associated with convective shutdown, although Section \ref{ssec:prof} showed that convective stability is preferred under reducing conditions for permanent magma oceans. The composition of the atmospheres could potentially be inferred from these emission spectra, and thereby allow inference on the convective regime of these atmospheres.

\subsection{Chemistry case study}
\label{ssec:chem}
\begin{figure}
    \centering
    \includegraphics[width=\linewidth, keepaspectratio]{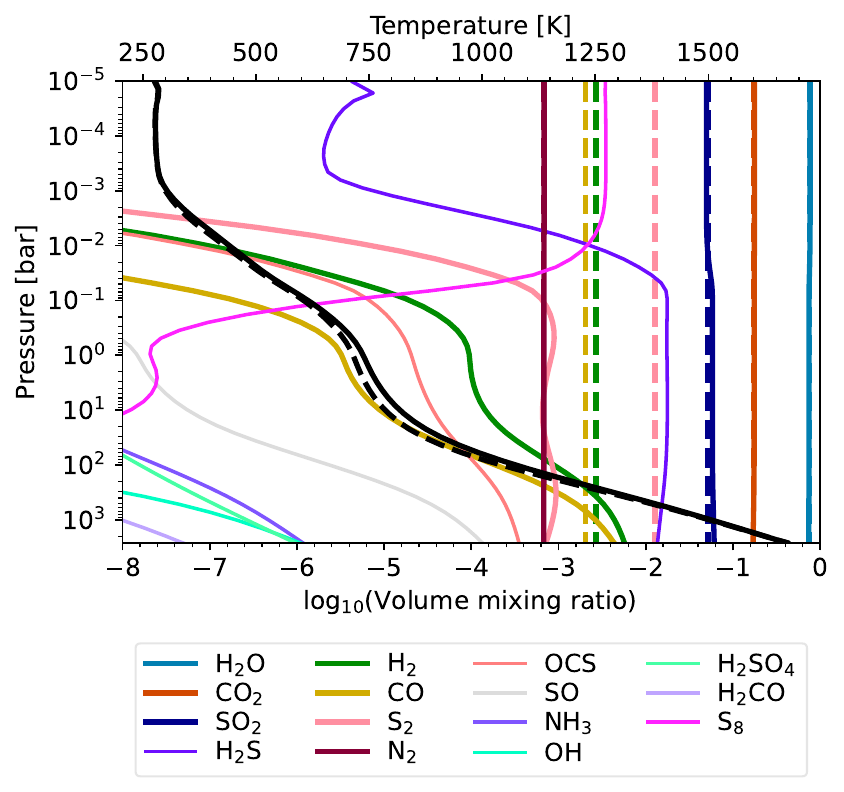}
    \caption{Composition and temperature versus pressure in the model atmosphere of a young TRAPPIST-1 c (IW+5). This corresponds to the point of magma ocean solidification, with elemental abundances derived from time-evolution with AGNI/PROTEUS (dashed lines) and then used to determine gas volume mixing ratios re-calculated self-consistently with AGNI/FastChem (solid lines). The coloured lines plot volume mixing ratios (bottom x-axis). The black lines plot temperature (top x-axis). Species with volume mixing ratios $< 10^{-8}$ are not plotted. The legend is sorted by surface mixing ratios in descending order.}
    \label{fig:chem}
\end{figure}

In this section we couple FastChem and AGNI in order to probe how equilibrium chemistry impacts the composition of the atmospheres on recently solidified planets. This neglects the role of diffusive processes on setting the composition of the atmosphere. However, the results presented in Section \ref{ssec:prof} have shown that it is possible for the atmospheres on these planets to be stable to convection, which means that the atmospheric composition will be controlled primarily by chemical processes rather than by dynamics. For the case of TRAPPIST-1 c at IW+5, which is shown to be entirely stable to convection (Fig. \ref{fig:prof}), we use the outgassed elemental composition and $T_{\text{m}}$ at the point of solidification (Fig. \ref{fig:pies}) to re-calculate the gas-phase speciation and temperature structure self-consistently. Simulations with PROTEUS (Fig. \ref{fig:melt}) indicate that the magma ocean solidifies in this case, so volatile dissolution plays a negligible role in setting the atmospheric composition. The new atmospheric composition is plotted with solid lines in Fig. \ref{fig:chem} as volume mixing ratio (x-axis) versus total pressure (y-axis). The composition determined from the time-evolved calculation, which assumes that the atmosphere is isochemical, is indicated with the dashed lines. The corresponding temperature profiles are plotted in black

The maximum relative difference is temperature is \SI{38.75}{\kelvin} (4.36 per cent) between the original (dashed black line) and self-consistent (solid black line) calculations. This corresponds to a pressure level of approximately \SI{36}{\bar}.  The difference in outgoing longwave radiation is \SI{0.02}{\WPMS}, which is small because the temperature profiles and abundances of three major gases (\ce{H2O}, \ce{CO2}, \ce{SO2}) are effectively identical between the two models in the upper atmosphere. The maximum relative difference in the atmospheric mean molecular weight is 1.10 per cent between the two models. 
\par 
The calculations with AGNI/FastChem (solid lines) show that the abundances of \ce{H2} (green line) and \ce{CO} (yellow line) decline to less than $10^{-8}$ in the upper atmosphere. The hydrogen atoms are instead favourably speciated into \ce{H2O}, and condensation remains absent. Production of \ce{S8} (magenta line) is also predicted in this region (magenta line), with a corresponding minor decrease in the abundance of \ce{SO2}.

\section{Discussion}
\label{sec:discuss}

\subsection{Convective stability in magma ocean atmospheres}
Evolution models of HD 63433 d show that it is possible for atmospheres of mixed volatile composition overlying magma oceans to be stable to convection (Fig. \ref{fig:prof}). This fits with the predictions of \citet{nicholls_proteus_2025} based on radiative heating rates in mixed atmospheres, and with those of \citet{selsis_cool_2023} for pure-steam atmospheres. The transport of energy throughout the atmosphere column becomes entirely handled by radiation and sensible heating when convection shuts down. Convective stability occurs in the model atmospheres of HD 63433 d when radiative transport is sufficient to carry the approximately zero net outgoing energy flux $F_t$ that exists at radiative equilibrium. This small $F_t$ does not itself induce temperature gradients sufficient for triggering convection. The square scatter points in the top panel of Fig. \ref{fig:prof} show that oxidising cases maintain convection in the middle atmosphere ($p < \SI{1}{\bar}$), which is triggered by the absorption of downwelling shortwave radiation from the host star and not by transport of heat from a cooling interior. In these regions, convection transports energy upwards to offset the net downward transport of energy by radiation (c.f. the middle panel of Fig. \ref{fig:valid}). The reducing cases of HD 63433 d are fully radiative in part due to backscattering of downwelling stellar radiation enabled by the large Rayleigh scattering cross-sections of \ce{H2} and \ce{CO} \citep{pierrehumbert_book_2010}. While the surface temperatures predicted by AGNI are $\sim \SI{2000}{\kelvin}$ lower than those predicted by JANUS, the upper-atmosphere temperatures are up to \SI{1000}{\kelvin} larger due to radiative heating by absorption of stellar radiation. This means that, despite the shallower lapse rate (i.e. smaller $|\di \ln T / \di \ln P|$) associated with convective stability in these atmospheres, the surface temperatures remain sufficient to sustain a permanent magma ocean (Fig. \ref{fig:melt}). This shows that convective shutdown does not necessarily preclude a permanent magma ocean on terrestrial-mass exoplanets, and that the findings of \citet{selsis_cool_2023} with \ce{H2O}-dominated atmospheres cannot be broadly extended to planetary evolution. It is also important to note that \ce{H2O}-dominated atmospheres are not generated by any of the models of HD 63433 d. The simulated composition of HD 63433 d's atmosphere (Fig. \ref{fig:pies}) does not vary significantly between the AGNI and JANUS cases, as they both maintain significant mantle melt fractions (Fig. \ref{fig:melt}) at high surface temperatures. The composition varies with the surface oxygen fugacity across the same three regimes shown in Fig. \ref{fig:outgas}.
\par 

Time-evolved simulations of a young TRAPPIST-1 c with AGNI never reach radiative equilibrium and instead cool until the magma ocean solidifies. The different behaviour between the two atmosphere models is in line with the predictions of \citet{selsis_cool_2023}. We do not simulate the post-solidification evolution of TRAPPIST-1 c. These models of TRAPPIST-1 c all have a non-zero net upward heat fluxes $F_t$ -- equivalent to a non-zero $T_{\text{int}}$ -- at the point of solidification: \SI{4.9}{\WPMS} at IW-5, decreasing monotonically to \SI{0.6}{\WPMS} at IW+5. This corresponds with convective instability in these models (square scatter points in the bottom panel of Fig. \ref{fig:prof}), where convection is maintained under reducing conditions with larger $F_t$, while the atmospheres become purely radiative under oxidising conditions with smaller $F_t$. This is opposite to the trend seen for HD 63433 d, as convection in the atmosphere of TRAPPIST-1 c is triggered primarily by upward transport of energy from the cooling magma ocean, rather than by the absorption of downwelling stellar radiation. Additionally, the lower photospheric temperature of the host star and smaller instellation of TRAPPIST-1 c also means that the amount of optical radiation readily available to trigger convection is reduced compared to HD 63433 d. The temperature structures of the reducing TRAPPIST-1 c cases effectively reproduce the canonical model for the structure of these atmospheres: a deep dry troposphere and an approximately-isothermal stratosphere, although these atmospheres remain free of condensation in all cases \citep{hamano_lifetime_2015, lichtenberg_redox_2021, nicholls_proteus_2025}. Further time-evolution and cooling of the surface may see the deep isothermal layers form, as the planet approaches radiative equilibrium. The stratospheric temperature varies between $\sim 300$ and $\SI{500}{\kelvin}$. The near-isothermal stratosphere formed under reducing conditions has a temperature consistent with analytical estimates of the planet's equilibrium temperature (\SI{508}{\kelvin}) when factoring in the relatively low Bond albedo (determined by SOCRATES as $\sim 1 \text{per cent}$) and the increased stellar luminosity at this early stellar age \citep{spada_radius_2013, baraffe_new_2015}. The TRAPPIST-1 c cases simulated with JANUS maintain high surface temperatures with nearly 100 per cent melt fractions (Fig. \ref{fig:melt}). The stark contrast between the JANUS and AGNI simulation outcomes demonstrates the necessity of using more realistic models of energy transport within their atmospheres.
\par

\subsection{Observational prospects for HD 63433 d and TRAPPIST-1 c}
The infrared emission spectra plotted in the top panel of Fig. \ref{fig:emit} indicate that an observational \mbox{(non-)detection} of atmospheric \ce{CO2} and \ce{SO2} may be used to constrain the surface oxygen fugacity of lava worlds (e.g. HD 63433 d), as absorption features associated with these gases  vary strongly with mantle $f\ce{O2}$. Since HD 63433 d is only 414 Myr old, it may be possible that it still maintains a permanent magma ocean beneath an outgassed volatile atmosphere with a composition set by equilibrium chemistry and dissolution into the melt. This would allow observations of HD 63433 d to be used to place constraints on the redox state of the magma ocean. There is evidence that Earth's upper mantle has remained oxidised since at least 3.48 Ga, and was oxidised within 200 Myr of formation \citep{trail_hadean_2011, nicklas_redox_2018}. Detection of \ce{SO2} or \ce{CO2} within the atmosphere of HD 63433 d would then imply a similar redox history to Earth's. In particular, emission observations of HD 63433 d with MIRI F770W could potentially constrain the brightness within this \ce{SO2} band, and thereby the oxidation state of an underlying magma ocean. Attempting to constrain atmospheric composition with a single photometric point gives rise to various degeneracies; Fig. \ref{fig:emit} shows that MIRI LRS and NIRISS SOSS simultaneously span multiple features which are sensitive to $f\ce{O2}$, so observations with these instruments could alternatively provide deeper insight into these atmospheres compared to photometry \citep{hammond_reliable_2024, zieba_no_2023, august_hot_2024}. Given the wide range of outgassed atmospheric compositions presented in Fig. \ref{fig:pies}, high-resolution observations of HD 63433 d -- such as with the upcoming ELT -- may be able to probe abundance ratios of gases in the atmosphere of HD 63433 d. HD 63433 d is an optimal target for future direct imaging missions, such as the LIFE mission, which will aim to discern the atmospheric composition of terrestrial protoplanets to gain insights into the climate state of Hadean-like surface environments \citep{Bonati2019,Cesario2024}.
\par

The bottom panel of Fig. \ref{fig:emit} shows the modelled emission spectrum from TRAPPIST-1 c at the point of magma ocean solidification. The absorption features are comparable with the top panel for the most oxidising cases (>IW+2). However, the near-isothermal upper atmosphere generated by a reducing magma ocean on this planet (c.f. bottom panel of Fig. \ref{fig:prof}) yields a near-blackbody emission spectrum with comparatively small absorption features. The outgoing radiation continuum for $f\ce{O2} \le IW+1$ closely matches the blackbody curves, which corresponds to the isothermal stratospheres formed by these models. The oxidising cases yield stratospheric temperatures of approximately \SI{300}{\kelvin} (Fig. \ref{fig:prof}), which corresponds to the $\sim \SI{300}{\kelvin}$ brightness temperature in the \ce{CO2} and \ce{SO2} absorption bands in the emission spectra at $f\ce{O2} > IW+1$. It is important to note that these spectra represent the planet-averaged outgoing radiation, while secondary eclipse observations \citep{zieba_no_2023} probe the day-side atmosphere and yield relatively higher brightness temperatures, subject to redistribution of heat by atmospheric dynamics \citep{pierrehumbert_atmospheric_2019, hammond_rotational_2021}. The extended age of the TRAPPIST-1 system compared to the HD 63433 system means that the planets orbiting TRAPPIST-1 were/are subject to several physical processes not modelled in this work, such as atmospheric escape, which could make direct comparisons between these models and future JWST observations of these planets difficult. Atmospheres on TRAPPIST-1 c could be entirely removed by escape processes, or at least strongly fractionated by the escape of low molecular weight species \citep{owen_review_2019}. Additionally, the composition of any remaining atmosphere could be influenced ongoing volcanic outgassing enabled by radiogenic or tidal heating \citep{matsuyama_io_2022, tyler_tidal_2015, seligman_potential_2024}. Future theoretical studies should consider using the outcomes of coupled interior-atmosphere models as the starting point for further secular evolution \citep{krissansen_erosion_2024}. The presence of surface liquid water could enable the geochemical cycling of volatiles (e.g. \ce{CO2}), which would exert control over the compositions of their atmospheres after their magma oceans have solidified \citep{hakim_weathering_2021, graham_weathering_2024}. The physical processes modelled in this work are not unique to TRAPPIST-1 c and HD 63433 d, so it is quite possible that the results of these simulations may be consistent with future observations of younger exoplanets with masses and host stars similar to TRAPPIST-1 c and HD 63433 d \citep{diamond_hot_2023, scarsdale_compass_2024, alam_compass_2024}.
\par 

\subsection{Compositional inhomogeneity of atmospheres at chemical equilibrium}
We have generally assumed that these atmospheres can be modelled with well-mixed compositions set by outgassing from an underlying magma ocean. This assumption is consistent with previous work \citep{elkins_linked_2008, hamano_lifetime_2015, nicholls_proteus_2025}. However, the results presented in Section \ref{ssec:chem} indicate that gas-phase equilibrium chemistry may be sufficient to generate compositional inhomogeneity throughout the atmosphere. This is entirely due to variation in the local pressure and temperature. The three most abundant gases produced by the outgassing model (\ce{H2O}, \ce{CO2}, \ce{SO2}) remain approximately isochemical when the speciation is re-calculated by FastChem. However, \ce{H2} and \ce{CO} are found to have significantly reduced abundances in the upper atmosphere. This has a negligible impact on the energy balance of the atmosphere (outgoing longwave radiation differs by \SI{0.02}{\WPMS} between the models) because the abundances differ only in regions of low pressure, which are optically thin. At the surface, the most abundant species unmodelled in the main evolutionary simulations is \ce{H2S}. The opacity of \ce{H2S} is included in the radiative transfer calculations, but the small change in outgoing radiation associated with the self-consistent chemistry model shows that it has a very minor impact on the energy balance of the atmosphere. However, \ce{H2S} could potentially be observable in infrared emission spectra alongside \ce{SO2} and \ce{S2} \citep{janssen_sulfur_2024}. The formation of \ce{S8} in the upper atmosphere -- through photolysis of \ce{H2S} \citep{tsai_vulcan_2021} and thermochemical production (Fig. \ref{fig:chem}) -- could lead to the production of hazes which dampen absorption features generated deeper in the atmosphere and complicate the characterisation of these planets from observations \citep{gao_sulfur_2017,pavlov_anoxic_2002,toon_the_1982}. 
\par 

Convection could act as an important mixing process in these atmospheres, but our modelling under the Schwarzschild criterion for convective instability suggests that convection does not always occur (Fig. \ref{fig:prof}). Convection could additionally be inhibited by compositional gradients in low molecular weight atmospheres \citep{habib_convection_2024, innes_runaway_2023}. This would correspond to atmospheres formed in equilibrium with low-$f\ce{O2}$ melt, or those with large gas envelopes captured from the stellar nebula / protoplanetary disk, or hot atmospheres containing rock vapours \citep{van_buchem_lavatmos_2023, zilinskas_observability_2023}. In the absence of strong atmospheric mixing, (photo)chemistry in the upper atmosphere is expected to stratify the atmospheric composition and complicate inferences of the conditions in the deeper atmosphere. Future work should explore the location of the quench point in these atmospheres under various mixing scenarios with the view of understanding its impact on observables. This may require hydrodynamic models to resolve the convection \citep{habib_convection_2024}, or global circulation models to resolve day-night differences. In addition, the exact composition of the upper atmosphere will influence the rate at which it can escape to space \citep{owen_review_2019}.
\par

\subsection{Limitations and future work}
Introduction of additional interior heating processes, such a tidal heating, could allow for permanent magma oceans in cases which would otherwise solidify. Energy dissipated in the mantles of these planets would ultimately have to be transported upwards through their atmospheres and into space. For TRAPPIST-1 c, this could stall further cooling and prolong convection in its atmosphere. Although we do not model tidal heating in this work, numerical simulations of tides in the TRAPPIST-1 system suggest that planet-star tides can dissipate significant amounts of energy into their interiors \citep{hay_tides_2019}. The maximal estimates for the power dissipated in TRAPPIST-1 c by \citet{hay_tides_2019} are $\sim \SI{1e18}{\watt}$. Globally averaged, this corresponds to a heat flux $\sim \SI{1000}{\WPMS}$. Future work should investigate whether tidal heat dissipation can maintain atmospheric convection and/or a permanent magma oceans on the planets orbiting TRAPPIST-1. Similarly, the terrestrial exoplanet LP 98-59 d exists within a system of four planets and has been confirmed to host an atmosphere containing \ce{SO2} and \ce{H2S}; this may be consistent with a permanent magma ocean enabled by planet-planet tidal interactions \citep{banerjee_atmospheric_2024}.
\par 

We have applied the semi-empirical MT\_CKD model to simulate the continuum absorption of \ce{H2O} \citep{mlawer_mtckd_2012, mlawer_mtckd_2023}. Although MT\_CKD is only calibrated at temperate conditions, this has been a common assumption throughout the literature for lack of alternative prescriptions \citep{hamano_lifetime_2015, selsis_cool_2023, boukrouche_beyond_2021, nicholls_proteus_2025}. Application of MT\_CKD to the study of atmospheres containing large inventories of \ce{H2O} could be a source of error in the calculations of their opacity, and therefore in the cooling rate of these planets. This motivates future studies with physical experiments or molecular dynamics models on the collisional continua of \ce{H2O} and \ce{H2} at high temperatures and pressures.
\par 

Oxygen fugacity plays a key role in setting the outcome of magma ocean evolution through its control over atmospheric composition \citep{nicholls_proteus_2025, sossi_redox_2020}. We have shown in this work that these variable compositions influence the presence of atmospheric convection (Section \ref{ssec:prof}) and also generate distinct absorption features in emission spectra (Section \ref{ssec:emit}). In reality, magma ocean $f\ce{O2}$ is not a conserved quantity and would instead vary in time if the redox state of \ce{Fe} (as well as \ce{Cr} or \ce{C}) in the magma evolves as it reacts with \ce{H2O} or \ce{H2} \citep{HIRSCHMANN_2022, Itcovitz_2022, krissansen_erosion_2024}. Our results can still be applied when considering that a range of magma ocean surface $f\ce{O2}$ are expected as an outcome of internal evolution, for example due to varying core formation conditions \citep{guimond_mineralogical_2023, lichtenberg_review_2025, Deng2020}. Future work should study the influence of evolving $f\ce{O2}$ during magma ocean cooling and crystallisation.
\par 

Atmospheric escape may play a role in setting the evolution of these planets, as it removes the blanketing introduced by atmospheric opacity. Models have shown that thermal radiation from the deep atmosphere alone may be sufficient to drive the escape of the upper atmosphere \citep{guo_escape_2024}. This may be particularly important for older planets as escape can continue to take place long after a magma ocean has solidified. Compositional fractionation through the preferential escape of lighter elements would also influence atmospheric composition and act to oxidise the atmosphere and magma ocean over time \citep{hunten_escape_1987, kaye_escape_1987, zahnle_escape_2023}. Under reducing conditions with AGNI, the TRAPPIST-1 c models support \ce{H2}-\ce{CH4} dominated atmospheres at the point of solidification. The hydrogen in these species would be preferentially lost to space through escape processes, which could potentially yield a thin carbon-rich atmosphere and/or eventually yield a sooty bare-rock surface. However, under oxidising conditions the atmosphere is instead \ce{H2O}-\ce{CO2} dominated (\SI{459.3}{\bar} of \ce{CO2} at IW+5), which could prove more difficult to escape than the reducing atmosphere. Recent hydrodynamic modelling of \ce{H2}-rich atmospheres on Earth-mass planets has shown that thermal cooling by dilute \ce{CO2} and \ce{CO} can reduce the efficiency of atmospheric escape by up to an order of magnitude \citep{yoshida_escape_2024}. Regardless, the influence of atmospheric escape does not entirely prevent the association of observed atmospheric absorption features with particular mantle geochemistry -- assuming that the atmosphere and magma ocean remain near equilibrium -- but it would require models to simultaneously account for escape processes and time-variable $f\ce{O2}$. Observational indications of a reducing atmosphere (e.g. one with a significant abundance of \ce{CO}) could also imply incomplete core segregation \citep{lichtenberg_review_2025}. It is also worth noting that, because HD 63433 is a relatively high-mass star, its X-ray saturation timescale would have been shorter than that of TRAPPIST-1 \citep{johnstone_active_2021}. This means that the time-integrated exposure of HD 63433 d to the high-energy photons necessary to drive atmospheric escape would have been comparatively small, thereby potentially allowing the planet to retain an atmosphere to this day \citep{owen_review_2019, yoshida_escape_2024}.
\par 

Observations by \citet{hu_55cnc_2024} of the ultra short period super-Earth 55 Cancri e have suggested the presence of a volatile-rich atmosphere, with subsequent re-visits by JWST showing substantial variation in the derived spectrum \citep{patel_55cnc_2024, loftus_55cnc_2024}. It is likely that this planet maintains a permanent magma ocean due to intense stellar heating. Detection of \ce{SO2} in its atmosphere, or constraints on the relative abundances of \ce{CO2} and \ce{CO} (see trends in Figures \ref{fig:outgas} and \ref{fig:pies}), would indicate that 55 Cancri e currently has an oxidised interior through the same analysis applied to the synthetic spectra of HD 63433 d (Fig. \ref{fig:emit}). While the 8 Gyr age of 55 Cancri e likely provides ample time for escape processes to strip the planet of volatiles, it is possible that its current atmosphere could be supplied by continued outgassing from its interior, which buffers the remaining volatile reservoir against escape processes \citep{Meier2023,Heng2023}.
\par 

Future observations of TRAPPIST-1 c and HD 63433 d may be able to provide further constraints on their atmospheres and composition. \citet{zieba_no_2023} used a single photometric point to infer that TRAPPIST-1 c may have only a tenuous atmosphere, but phase curve observations would more directly probe the (lack of) heat redistribution occurring on the planet \citep{hammond_reliable_2024}. The small transit depth ($\sim 100$ ppm) and correspondingly small radius of HD 63433 d measured by \citet{capistrant_hd63433d_2024} could indicate that HD 63433 d has already lost its atmosphere or is dominated by high-mean molar mass gases. This could be tested by radial-velocity measurements of the mass of HD 63433 d, which would allow for estimations of its bulk density. Phase curve observations may further constrain the planets' day-side temperatures and/or heat redistribution efficiency \citep{koll_scaling_2022, hammond_reliable_2024}. If future observations indeed reveal HD 63433 d to be a bare rock, despite our modelling suggesting the presence of permanent magma ocean, this would indicate that efficient escape processes have acted to strip the planet of its atmosphere despite the buffering of volatiles by dissolution into the underlying melt \citep{owen_review_2019, zahnle_escape_2023}. Confirmation of a volatile atmosphere on HD 63433 d would suggest that the atmospheres of rocky exoplanets can be buffered against escape by volatile dissolution into magma oceans \citep{DornLichtenberg2021,lichtenberg_review_2025}.
\par

\section{Conclusions}
\label{sec:conclude}

With the aim of investigating convective stability in the atmospheres overlying magma oceans, we have extended upon the PROTEUS framework described by \citet{nicholls_proteus_2025} with the introduction of a new radiative-convective atmosphere model. We use this framework to simulate the early evolution of the exoplanets HD 63433 d and TRAPPIST-1 c. Our conclusions are as follows.
\begin{enumerate}
    \item Atmospheres generated by magma ocean degassing can be stable to convection, depending on their gaseous composition and the spectrum of incoming stellar radiation. We have shown that even with the shutdown of deep convection, it is possible to maintain a permanent magma ocean for some planets, which means that the predictions of \citet{selsis_cool_2023} cannot be directly extended to atmospheres of mixed composition. Reducing atmospheres (abundant in low mass species such as \ce{H2} and \ce{CO}) overlying permanent magma oceans are likely to be entirely stable to convection, while oxidising atmospheres can maintain convection far above their surfaces triggered by absorption of stellar radiation. It is also possible to trigger atmospheric convection from the surface of a planet upwards by the release of heat from its interior.
    
    \item Time-evolved models indicate that HD 63433 d may maintain a permanent magma ocean to this day. The amount of melt within the interior is strongly linked to the composition of the overlying atmosphere, which introduces a physical feedback that can only be handled by time-integrated modelling. Future observations of this planet's infrared emission spectrum would allow for the characterisation of its atmosphere, which may constrain the redox state of its mantle through the (non-)detection of \ce{CO2} and \ce{SO2}.

    \item Reduced \ce{H2}-dominated atmospheres may exhibit isothermal temperature profiles which generate relatively featureless emission spectra that are well-matched by a blackbody. This indicates that secondary-eclipse photometry may not always be a reliable discriminator between planets with thin or non-existent atmospheres versus those with isothermal profiles that extend into optically thick regions.
\end{enumerate}

Future work should consider the role of mixing and chemical processes in relating upper-atmospheric composition to that at the surface in equilibrium with an underlying magma ocean.

\section*{Acknowledgements}
CRediT author statements. \textbf{HN}: Methodology, Software, Formal analysis, Investigation, Validation, Writing - Original Draft, Writing - Review \& Editing, Visualization. \textbf{RTP}: Conceptualization, Supervision, Writing - Review \& Editing. \textbf{TL}: Conceptualization, Writing - Review \& Editing, Project administration. \textbf{LS}: Methodology, Software, Data Curation. \textbf{SS}: Methodology, Software, Data Curation.
\par 
%{\color{magenta} % revision 1
HN thanks Claire Guimond and Tad Komacek for their thoughtful inputs. TL acknowledges support from the Netherlands eScience Center under grant number NLESC.OEC.2023.017, the Branco Weiss Foundation, the Alfred P. Sloan Foundation (AEThER project, G202114194), and NASA’s Nexus for Exoplanet System Science research coordination network (Alien Earths project, 80NSSC21K0593). RTP acknowledges support from the UK STFC and the AEThER project. 
\par 
We thank our anonymous reviewer for their positive and constructive feedback which expanded the implications of this work. 
%}

%%%%%%%%%%%%%%%%%%%%%%%%%%%%%%%%%%%%%%%%%%%%%%%%%%
\section*{Data Availability}
PROTEUS\footnote{\url{https://github.com/FormingWorlds/PROTEUS}} and AGNI\footnote{\url{https://github.com/nichollsh/AGNI}} are open source software available on GitHub. The data underlying this article are available on Zenodo: \citet{this_data}.

\bibliographystyle{mnras}
\bibliography{main.bib} 

%%%%%%%%%%%%%%%%% APPENDICES %%%%%%%%%%%%%%%%%%%%%

%{\color{magenta} % revision 1

\appendix

\section{JANUS emission spectra}
\label{app:emit_janus}
The character of a planet's emission spectrum is primarily determined by its atmospheric composition and structure, and the irradiation that it receives from its star \citep{perryman_imaging_2011}. It was shown in Section \ref{ssec:prof} that the temperature profiles modelled by JANUS differ significantly compared to those modelled by AGNI, especially with respect to the formation of radiative layers in the latter cases. In Fig. \ref{fig:emit_janus} below, we additionally present the emission spectra corresponding to the cases simulated with JANUS. These have been calculated at an increased resolution compared to the main simulations, and are analogous to those presented in Fig. \ref{fig:emit} for simulations with AGNI.
\par 
The absorption features in these alternative emission spectra of HD 63433 d (top panel of Fig. \ref{fig:emit_janus}) are common to those in Fig. \ref{fig:emit}. This follows from the fact that the composition is calculated in the same manner, although generally at higher surface temperatures. In contrast to the AGNI cases, the temperature profiles are less sensitive to the composition of the atmosphere because they are set only by an analytical moist pseudoadiabat \citep{graham_multispecies_2021}; this can be clearly seen by the dashed lines in Fig. \ref{fig:prof}. The emission spectra presented in Fig. \ref{fig:emit_janus} depend less strongly on $f\ce{O2}$ as a result of this partial decoupling between $T(p)$ and $f\ce{O2}$. This indicates that the application of sufficiently comprehensive atmosphere models (e.g. AGNI) can be necessary for calculating meaningful emission spectra. Similar findings also exist within the literature (e.g. \citet{ drummond_effects_2016, malik_helios_2017, nicholls_temperaturechemistry_2023}).
\par 
The emission spectra resulting from simulations of these two planets with JANUS are relatively similar to each other: compare the top and bottom panels of Fig. \ref{fig:emit_janus}. This is also a result of the moist pseudoadiabat prescription within JANUS, which cannot capture key characteristics of the atmospheric temperature. For example, the formation of a near-isothermal stratospheres under reducing conditions on TRAPPIST-1 c (Fig. \ref{fig:prof}).

\begin{figure*}
    \centering
    \includegraphics[width=0.93\linewidth, keepaspectratio]{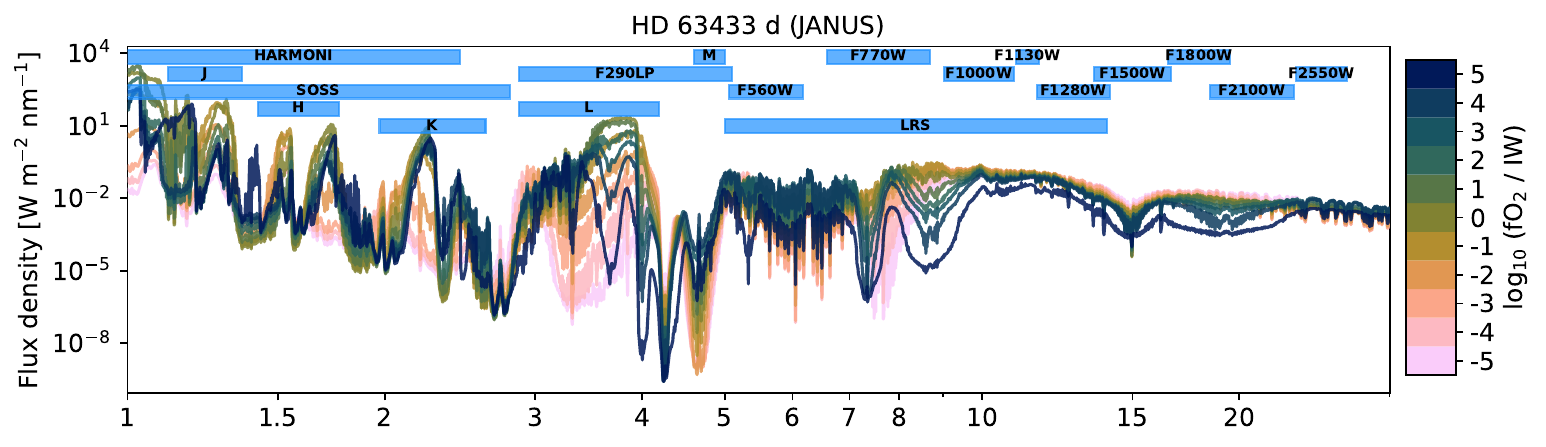}
    \\
    \vspace{-2mm}
    \includegraphics[width=0.93\linewidth, keepaspectratio]{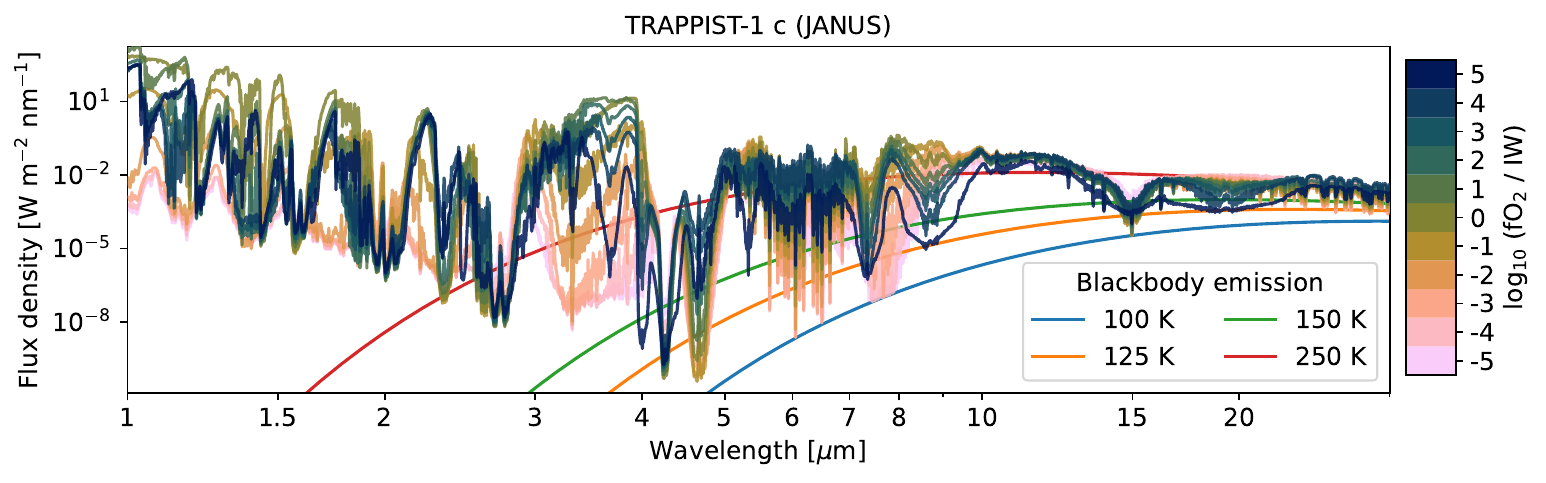}%
    \caption{Top-of-atmosphere flux representative of the planet-averaged outgoing radiation from simulations of young HD 63433 d and TRAPPIST-1 c. This is the same as Fig. \ref{fig:emit}, but instead corresponds to the results of simulations coupled to JANUS. All of these cases are at radiative equilibrium and do not solidify within these simulations.}
    \label{fig:emit_janus}
\end{figure*}

%}

%%%%%%%%%%%%%%%%%%%%%%%%%%%%%%%%%%%%%%%%%%%%%%%%%%
% Don't change these lines
\bsp	% typesetting comment
\label{lastpage}
\end{document}